\documentclass[aps,prb,twocolumn,floatfix,showkeys,superscriptaddress,amssymb,preprintnumbers]{revtex4-1}

\usepackage[latin1,utf8]{inputenc}  
\usepackage[sc]{mathpazo}
\usepackage[T1]{fontenc}       
\usepackage[english]{babel}    
\usepackage{fancyhdr} 
\usepackage[dvips]{graphicx}
\usepackage{floatflt,epsfig}
\usepackage{float} 
\usepackage{wrapfig}
\usepackage{setspace}

\usepackage{amsmath}
\usepackage{amssymb}
\usepackage{amsfonts}
\usepackage{amsthm}
\usepackage{pictexwd,dcpic}
\usepackage{amscd}
\usepackage{color}
\usepackage{rotating}
\usepackage{hyperref}
\usepackage{cancel}
\usepackage{mathcomp}
\usepackage{xcolor}
\usepackage{times}

\begin{document}
	
	\title{Concentration-dependent atomic mobilities in FCC CoCrFeMnNi high-entropy alloys}
	
	\author{Daniel Gaertner}\email{For correspondence: daniel.gaertner@wwu.de}
	\affiliation{Institute of Materials Physics, University of M\"unster, D-48149 M\"unster, Germany}
	\author{Katrin Abrahams}
	\affiliation{ICAMS, Ruhr University Bochum, Universitätsstr. 150, 44801 Bochum, Germany}
	\author{Josua Kottke}
	\affiliation{Institute of Materials Physics, University of M\"unster, D-48149 M\"unster, Germany}
	\author{Vladimir A. Esin}
	\affiliation{MINES ParisTech, PSL Research University, Centre des Mat\'eriaux (CNRS UMR 7633), \'Evry, France}
	\author{Ingo Steinbach}
	\affiliation{ICAMS, Ruhr University Bochum, Universitätsstr. 150, 44801 Bochum, Germany}
	\author{Gerhard Wilde}
	\affiliation{Institute of Materials Physics, University of M\"unster, D-48149 M\"unster, Germany}
	\author{Sergiy V. Divinski}\email{For correspondence: divin@wwu.de}
	\affiliation{Institute of Materials Physics, University of M\"unster, D-48149 M\"unster, Germany}
	\date{\today}
	
	\begin{abstract}
		\noindent 
The diffusion kinetics in a CoCrFeMnNi high entropy alloy is investigated by a
combined radiotracer--interdiffusion experiment applied to a pseudo-binary
Co$_{15}$Cr$_{20}$Fe$_{20}$Mn$_{20}$Ni$_{25}$ /
Co$_{25}$Cr$_{20}$Fe$_{20}$Mn$_{20}$Ni$_{15}$ couple. As a result, the
composition-dependent tracer diffusion coefficients of Co, Cr, Fe and Mn are
determined. The elements are characterized by significantly different diffusion
rates, with Mn being the fastest element and Co being the slowest one. 
The elements having originally equiatomic concentration through the diffusion
couple are found to reveal up-hill diffusion, especially Cr and Mn. The atomic
mobility of Co seems to follow an S-shaped concentration
dependence along the diffusion path. The experimentally measured kinetic data are checked against the
existing CALPHAD-type databases. \\
In order to ensure a consistent treatment of tracer and chemical diffusion a
generalized symmetrized continuum approach for multi-component interdiffusion is
proposed.
Both, tracer and chemical diffusion concentration profiles are simulated and
compared to the measurements. By using the measured tracer diffusion
coefficients the chemical profiles can be described, almost perfectly, including
up-hill diffusion.
	\end{abstract}
	
	\maketitle

\section{Introduction}

In most of the engineering applications alloys are used which consist of one or
two element(s) as the principal element(s) and they are supplemented with (typically minor)
alloying elements to improve their physical and mechanical properties. However,
multi-principal-element alloys were not preferred, since according to the Gibbs phase
rule they lead potentially to formation of intermetallic compounds with usually
brittle complex structures. A new class of multicomponent alloys, called high
entropy alloys (HEAs), containing five or more principal elements in equiatomic
or nearly equiatomic proportions promise to provide attractive mechanical
properties including attractive strength-ductility combinations both at high-
and low temperatures \cite{Murty2014}. Due to their high configurational mixing
entropy ($\Delta S_{\text{mix}}$) HEAs were suggested to form fcc and/or bcc simple solid
solution phases instead of complex intermetallic phases \cite{Yeh2004}.\\
\noindent
As a counterpart to the configurational entropy, recent studies mention the importance of the
formation enthalpy in determining the phase
stability in HEAs. After Zhang et al. the high mixing entropy state does not
always have the lowest Gibbs free energy \cite{Zhang2014}. Moreover, complex
phases may precipitate in HEAs after long annealing treatments, typically at not
too high temperatures.
As well important as the configurational entropy are vibrational, electronic and
magnetic contributions to the entropy, shown by ab-initio calculations
for the CoCrFeMnNi alloy \cite{Ma2015}. Even short annealing of
the severely plastically deformed CoCrFeMnNi alloy at a temperature
of $450 ^{\circ}$C results in a phase decomposition, suggesting that a high
mixing entropy does not guarantee the phase stability \cite{Schuh2015,Otto2016}.
Furthermore, the single phase observed in HEAs might be a high temperature phase
with a kinetically constrained transformation \cite{Schuh2015}.\\
\noindent
Focusing on high temperature mechanical properties \cite{Guo2016,Chen2016}, creep strength
\cite{Lee2016,Zhang2016,Ma2016,Cao2016}, oxidation resistance
\cite{Kai2016,Laplanche2016,Holcomb2015} and coating applications
\cite{Shaginyan2016}, numerous HEAs have been investigated following an
originally introduced paradigm of four 'core' effects, i.e.
a high entropy, severe lattice distortion, 'cocktail' effect and 'sluggish' diffusion \cite{Yeh2004}. 
These basic principles are questioned now \cite{Pickering2016, cors},
nevertheless the understanding of the diffusion kinetics in HEAs, which is
assumed to be responsible for the unique features like excellent thermal
stability, decelerated grain growth, formation of nano-precipitates
\cite{Murty2014} and an excellent resistance to grain coarsening in a
nanocrystalline CoCrFeNi alloy \cite{Praveen2016}, is of fundamental
significance.\\
\noindent
The present knowledge about diffusion in HEAs is limited to few
interdiffusion investigations in couples or multiples \cite{Tsai2013,
Kulkarni2015, Dabrowa2016} and direct radiotracer diffusion measurements in polycrystalline
and single crystalline CoCrFeNi and CoCrFeMnNi \cite{Vaidya2016, Vaidya2017,
Vaidya2018, Gaertner2018}. Interdiffusion coefficients in a CoCrFeMn$_{0.5}$Ni
alloy were determined using a quasi-binary approach \cite{Tsai2013}, originally
known as pseudo-binary approach \cite{Paul2013}, proposing the evaluated diffusivities
to be approximately equal to the intrinsic and tracer diffusivities of the
equiatomic CoCrFeMnNi alloy with a thermodynamic factor of about unity
\cite{Tsai2013}. In fact, this assumption was found to be correct in the
framework of the random alloy model \cite{Murch2017}. However, the basic
principles of the analysis by Tsai et al. \cite{Tsai2013} were seriously
questioned recently \cite{Review}. The direct radiotracer measurements, being
focused on measuring the bulk and short-circuit diffusion rates of the
constituting elements in absence of any chemical interaction due to low
diffusant concentrations, are preferable but typically limited to single
compositions of the given alloy system. Moreover, recently it was shown that the
thermodynamic factor (more specifically the product of the thermodynamic factor
and the vacancy wind effect), being indeed about unity in CoCrFeNi, deviates
strongly from unity in CoCrFeMnNi \cite{Vaidya2018S}.\\
\noindent
In metallic materials the diffusion model implemented in the DICTRA (Diffusion
Controlled Phase Transformation) software is the most common continuum model
based on a sublattice description~\cite{Agren1982, Andersson1992,
Borgenstam2000}. In a three dimensional setting nowadays the multicomponent multiphase-field method with an integrated sublattice description is applied to phase transformations and microstructural evolution \cite{Zhang2015}.
In both implementations diffusion is combined with CALPHAD (Calculation of Phase Diagrams)
type thermodynamic and kinetic databases to account for temperature and
composition dependent Gibbs energies and atomic mobilities~\cite{Lukas2010,
Campbell2001}. The DICTRA diffusion model is based on a reference element, which is
predefined in most alloys by its principal element. In case of equiatomic
alloys, like HEAs, the selection of a reference element is arbitrary. Several
databases were developed for different main elements, e.g 
TCNI Ni-based Superalloys Database or Thermo-Calc Software TCFE Steels/Fe-alloys
Database. They can be extrapolated into the equiatomic region but this can lead
to inaccuracies. Currently thermodynamic databases especially designed for HEAs
were published:
Thermo-Calc Software TCHEA3 Database~\cite{TCHEA3} and another one developed by
Hallstedt's group~\cite{Haase2017}. Furthermore a mobility database
(Thermo-Calc Software MOBHEA1 Database~\cite{TCHEA_dat})
was published, which is based on the MOBNI4 mobility database~\cite{TCNI_dat}.\\\\
\noindent
The present work is focused on combined radiotracer and interdiffusion
experiments in HEAs determining the concentration dependent tracer diffusion
coefficients without estimation of the interdiffusion coefficients, that would
be conceptually hindered due to appearance of up-hill diffusion effects.
Simulations of both, the radiotracer and interdiffusion concentration profiles were
performed using a new generalized multi-component
diffusion model. This so-called pair-wise diffusion model (PD-model) is shown to
be especially appropriate for the compositions about equiatomic ones that makes
the model particularly suitable for HEAs.
In the binary case and in the dilute limit it reduces to the DICTRA model. The
simulations are used to compare the existing databases, with a special focus on multi-component diffusion kinetics
and cross correlation effects, with the newly determined composition dependent
tracer diffusion coefficients. The simulations show the importance of accurately
measured kinetic data combined with an appropriate diffusion model and
a database.

\section{Experimental procedure}

\subsection{Sample preparation}

Polycrystalline Co$_{15}$CrFeMnNi$_{25}$ and Co$_{25}$CrFeMnNi$_{15}$ samples
were produced by arc melting of a mixture of pure elements and homogenized
subsequently at $1473$ K for $48$ hours under purified Ar atmosphere. Here and
below the element concentrations are given in at.\% and, if not explicitly
specified by a proper sub-index, the element concentration is equal to 20~at.\%,
that corresponds to an equiatomic composition of the quinary alloy.\\
\noindent
Cylindrical samples with a diameter of $8$ mm and a thickness of $1.7$ mm
(Co$_{15}$CrFeMnNi$_{25}$) and $0.7$ mm (Co$_{25}$CrFeMnNi$_{15}$) were cut by
spark-erosion and etched carefully with aqua regia to remove any contamination.
The opposite faces of each specimen were polished by a standard metallographic
procedure to a mirror-like quality. A diffusion couple was assembled by fixing
the two samples and pressing them together by screws in a steel tube. Tungsten
discs were used as separators between the fixture and the samples. Two identical
couples -- one for the radiotracer and one for the interdiffusion experiments --
were prepared in order to prevent any radioactive contamination of the electron
probe microanalyzer (EPMA). The preparation of couples was performed in a
glove-box under a pure nitrogen atmosphere with $\approx1.5$ mbar
over-pressure.\\
\noindent
The assembled fixture was sealed into silica tubes under a purified ($5$N) Ar
atmosphere and subjected to the diffusion annealing at a temperature of $1373$ K
for $48$ hours. The temperature was measured and controlled by a Ni--NiCr
thermocouple to an accuracy of $\pm1$ K.

\subsection{Interdiffusion experiment}

After the diffusion annealing, one couple was embedded in epoxy and cut perpendicular to
the surface in two halves using a diamond wire saw. The halved disks were then
embedded in a conductive epoxy. Measurements using a CAMECA SX100 EPMA were
carried out at an accelerating voltage of $15$~kV and a beam current of $40$~nA
using pure standards for all elements. In order to measure the concentration
profiles of the constituting elements multiple dedicated line-scans
perpendicular to the interface between both sample parts were performed. The
line-scans were set to a total length of $400$ $\mu$m -- approximately $200$
$\mu$m in each sample half -- with a step size of $1$ $\mu$m.

\begin{figure*}
	\begin{center}
	\includegraphics[width=0.9\textwidth]{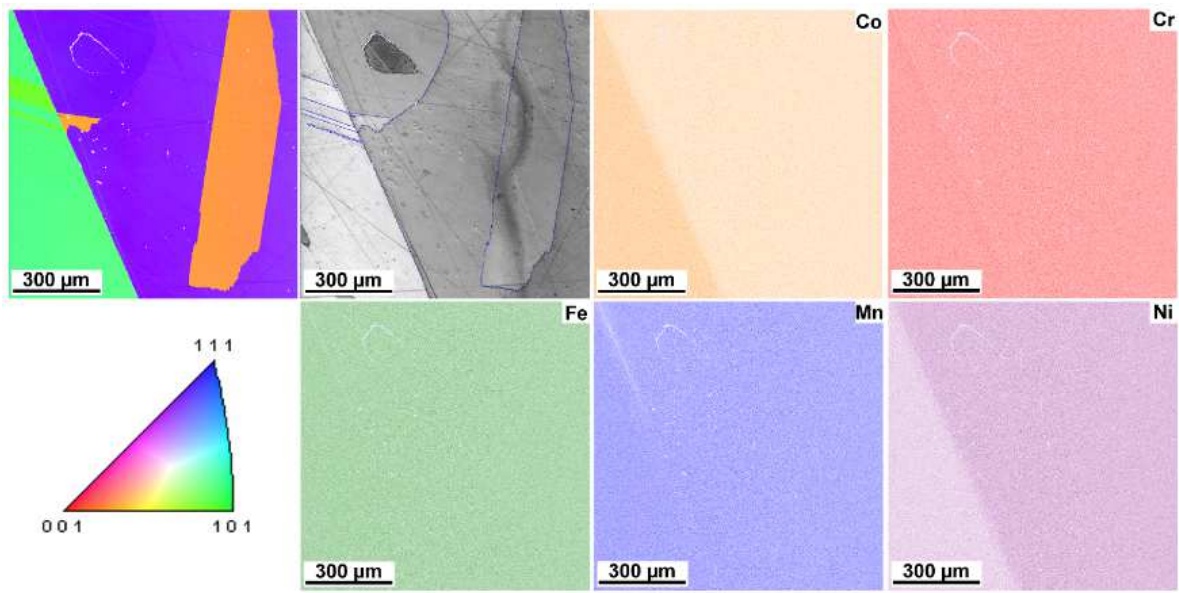}
	\end{center}
	\caption{\footnotesize Orientation imaging microscopy at the interface of the
		Co$_{25}$CrFeMnNi$_{15}$-Co$_{15}$CrFeMnNi$_{25}$ HEA couple and the
		corresponding elemental maps obtained by EDX analysis. The grain orientations
		are colored according to the inverse pole figure (bottom left panel).}
	\label{fig:HEAEBSD}
\end{figure*}

\subsection{Radiotracer experiment}

The radiotracers $^{57}$Co, $^{51}$Cr, $^{59}$Fe and $^{54}$Mn
were available as HCl solutions. The original solutions were highly diluted with
double-distilled water achieving the required specific activity of the tracer
material. A mixture of the tracers ($^{57}$Co$+^{51}$Cr$+^{59}$Fe$+^{54}$Mn) with
the radioactivity of about $5$ kBq for each tracer was applied on each polished
sample surface and dried. Subsequently the diffusion couple was assembled as
described above and subjected to the given diffusion annealing treatment. Since
all elements whose radioactive isotopes are used are already present in the
compound, their simultaneous application does not induce any additional
cross-correlation effects. Therefore, reliable data on tracer diffusion
coefficients in the alloys were obtained. Since the available $^{63}$Ni
radioisotope emmits only $\beta$-quanta, its decays cannot be recorded by the
$\gamma$-spectrometry. A separate (i.e. third) experiment would be required that is a subject for future work.\\
\noindent
After the diffusion annealing, the diffusion-bonded couple was reduced in diameter by about
$1.5$~mm in order to remove the effects of lateral and surface diffusion. The
penetration profiles were determined by precise parallel mechanical sectioning
using a grinding machine and grinding paper with SiC grains of about $30$
$\mu$m. Before and after sectioning the section masses were determined by
weighing the samples on a microbalance to an accuracy of $0.1\,\mu$g.\\
\noindent
The sectioning began from the Co$_{15}$CrFeMnNi$_{25}$ alloy (which was prepared
as a thicker disc) by gluing the Co$_{25}$CrFeMnNi$_{15}$ side to a holder. As
soon as the background for all isotopes was reached, the sectioning was stopped.
Then the couple was dismounted from the holder, reverted and glued again to the
holder by the Co$_{15}$CrFeMnNi$_{25}$ side. Afterwards, the sectioning was
continued from the Co$_{25}$CrFeMnNi$_{15}$ side till the Matano plane was
reached (that corresponded to an increase of the radioactivity) and then till
background was approached again. This approach allowed to measure in a single
experiment three concentration profiles: two profiles for tracer diffusion in
the unaffected end-members of the couple and one profile corresponding to tracer
diffusion in both directions from the Matano plane which proceeded parallel to
the chemical interdiffusion.\\
\noindent
A density variation induced in the alloy by chemical diffusion was neglected
that introduces an uncertainty in the depth coordinate $y$ below $1$~\%. Since
the initial thicknesses of the samples were carefully measured, a continuous
coordinate $y$ through the whole couple was in fact determined.\\
\noindent
The relative radioactivity of each section was measured by an available pure Ge
$\gamma$-detector equipped with a $16$ K multi-channel analyzer. All used
radioisotopes, $^{57}$Co, $^{51}$Cr, $^{59}$Fe and $^{54}$Mn, decay emitting the
$\gamma$-quanta whose energies \cite{Lemmer1955,Ofer1957,Heath1960,Lederer1978}
can easily be discriminated by the available setup with the energy resolution of
about $0.7$~eV. The relative radioactivities for each isotope were carefully
determined by the background subtraction, including the Compton scatter.\\
\noindent
The tracer concentration in a section is proportional to the section activity
divided by the section mass. As a result, the tracer concentration profiles,
$c^*_E(y)$, were determined where $E$ is the corresponding element, i.e.
$^{57}$Co, $^{51}$Cr, $^{59}$Fe or $^{54}$Mn.

\section{Experimental results}

\subsection{Microstructure analysis}

The microstructure and the chemical composition of the couple near the interface
after diffusion annealing was examined by orientation imaging microscopy
using Electron Back-Scatter Diffraction (EBSD) and Energy Dispersive X-Ray
Spectroscopy (EDX). Figure~\ref{fig:HEAEBSD} presents the region where
the original interface between both high-entropy alloys was located and shows
the grain orientation mapping using the inverse pole figure (bottom left) and the chemical maps. The grains were found to be
larger than $500$ $\mu$m on average and the chemical maps verify the homogeneity
of the equiatomic constituents in both alloys far from the Matano plane and
chemical gradients of Co and Ni at the interface. Furthermore, the chemical maps
reveal several local thin gaps between the two alloys (e.g. top left corner of
the given chemical maps). At such gaps the Mn concentration tends to be
decreased.
However, these spurious gaps are relatively small and localized and both alloys
are almost in perfect contact. At the positions with a perfect contact (and
simultaneously far from any grain boundary) the EPMA analysis was performed.
Correspondingly, composition profiles corresponding to true volume interdiffusion were determined.

\begin{figure*}[ht]
	\centering
	\includegraphics[width=0.9\textwidth]{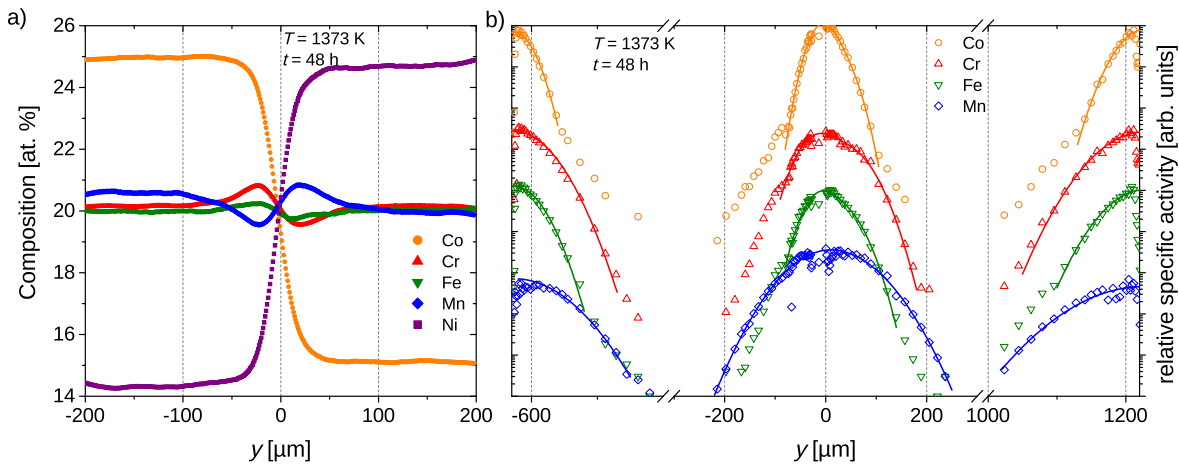}
	\caption{\footnotesize a) EPMA-analysis of the constituents at the interface
		and b) penetration profiles measured for tracer $^{57}$Co, $^{51}$Cr,
		$^{59}$Fe and $^{54}$Mn diffusion (open symbols) from the outer surfaces and
		the internal source located at the original interface between the two alloys
		(the Gaussian fits are represented by the straight lines). In b) the tracer
		profiles are shifted by multiplication with a constant factor for a better
		readability.}
	\label{fig:Diffusion}
\end{figure*}

\subsection{Diffusion experiments}

\subsubsection{EPMA interdiffusion measurements}

Figure~\ref{fig:Diffusion}a shows the concentration profiles of all
constituting elements measured by electron probe microanalysis. Each
profile was smoothed using the Savitzky-Golay filter method performing a local
second order polynomial regression over $50$ data points.\\
\noindent
The origin of the depth coordinate $y$ was set at the position of the Matano
plane \cite{Mehrer2007} of Co using

\begin{equation}
(c_{\text{L}}-c_{\text{R}})y_{\text{M}} +
\int_{c_{\text{L}}}^{c_{\text{M}}}ydc +
\int_{c_{\text{M}}}^{c_{\text{R}}}ydc = 0,
\end{equation}

\noindent with $c_{\text{L}}$ being the concentration on the left-hand side,
$c_{\text{R}}$ the concentration on the right-hand side, $c_{\text{M}}$ the
concentration at the Matano plane and $y_{\text{M}}$ the position of the Matano
plane. The position of the Matano plane was almost the same within the
experimental uncertainties when determined using the Ni concentration profile,
as it should be for a pseudo-binary couple \cite{Paul2013}. However, a careful
inspection of the concentration profiles in Fig.~\ref{fig:Diffusion}a reveals
that we are dealing with a non-ideal pseudo-binary couple (in terms of
Ref.~\cite{Review}) since the Ni and Mn concentrations in the nominally
Co$_{25}$CrFeMnNi$_{15}$ alloy deviate by less than $0.4$~at.\% from their nominal values.

\noindent
A remarkable feature is the appearance of up-hill diffusion in the
concentration profiles of the nominally equiatomic constituents Cr, Fe and Mn.
Especially, the Cr- and Mn-concentration profiles show distinct and oppositely
directed up-hill diffusion, Fig.~\ref{fig:Diffusion}a.\\
\noindent
The radiotracer experiment with combined interdiffusion was performed
separately under the same conditions like the sole interdiffusion experiment, at
$1373$ K for $48$ hours. The tracer solutions were applied on all four
polished sample surfaces. 

\subsubsection{Tracer diffusion measurements}

Figure~\ref{fig:Diffusion}b shows the measured penetration
profiles for tracer diffusion of $^{57}$Co, $^{51}$Cr, $^{59}$Fe and
$^{54}$Mn. The origin of the $y$ coordinate is set at the interface of the
diffusion couple which roughly corresponds to the Matano plane in
Fig.~\ref{fig:Diffusion}a. A slight disagreement between the $y$ scales in Figs.
\ref{fig:Diffusion}a and b stems from the accuracy of sample's thickness
measurements, (non-propagating!) error of thickness determination of individual
sections and the accuracy of the sample orientation for grinding perpendicular
to the diffusion direction.\\
\noindent
A comparison of Figs.~\ref{fig:Diffusion}a and b  substantiates that outer
tracer concentration profiles are located in regions without any influence of
the chemical driving force and correspond to diffusion in not-affected end-members
of the couple. Therefore, the corresponding tracer concentrations have to follow
a thin film solution of the diffusion problem \cite{Mehrer2007},

\begin{equation}\label{eq:Gauss}
c^*_E(y,t) = \frac{M_E}{\sqrt{\pi
D^*_E t}}\exp{\left(-\frac{(y-y_0)^2}{4D^*_E t}\right)}
\end{equation}

\noindent where $M_E$ denotes the initial amount of the tracer $E$, $c^*_E$
the concentration of the tracer $E$ in the layer, which is proportional to the
relative specific activity of the tracer, $y_0$ the origin of the diffusion
source, i.e. the left or right end of the couple, and $D^*_E$ the
corresponding volume diffusion coefficient. Excepting few very first data
points, all concentration profiles follow the Gaussian solutions over two to three orders of magnitude in
decrease of the tracer concentration, as indicated by the solid lines in
Fig.~\ref{fig:Diffusion}b.\\
\noindent
At larger depths, all penetration profiles -- both the outer concentration
profiles for end-members as well as the interface-related concentration profiles
-- reveal the existence of second, fast-diffusion branches. These branches
correspond to grain boundary diffusion in the polycrystalline alloys as it was
observed in our previous measurements of volume diffusion in CoCrFeNi and
CoCrFeMnNi alloys \cite{Vaidya2016, Vaidya2018}. In the present report we are
focused on the volume diffusion branches.\\
\noindent
The solid lines in Fig.~\ref{fig:Diffusion}b represent the expected Gaussian solutions of the first
diffusion branches which represent the true volume diffusion. From the fits, the tracer volume diffusion
coefficients, $D^*_E$, of all elements can be determined. The corresponding
parameters of the tracer diffusion experiments and the determined diffusion
coefficients in the unaffected end-members are summarized in Table
\ref{tab:OutDiffusion} for the Co$_{25}$CrFeMnNi$_{15}$ and Co$_{15}$CrFeMnNi$_{25}$ high-entropy alloys. For comparison, the tracer
diffusion coefficients measured for equiatomic CoCrFeMnNi alloys \cite{Gaertner2018} are given, too.

\begin{table}[ht]
\footnotesize
	\caption{\footnotesize Tracer volume diffusion coefficients $D^*$ (in
	$10^{-15}$~m$^2$s$^{-1}$) measured for the Co$_{25}$CrFeMnNi$_{15}$ and
	Co$_{15}$CrFeMnNi$_{25}$ high-entropy alloys at $1373$~K using unaffected
	end-members. The uncertainty of the $D^*$ values is typically below $20\,\%$.
	For comparison the averaged tracer diffusion coefficients determined for single
	crystalline equiatomic CoCrFeMnNi HEA \cite{Gaertner2018} are shown, too.}
	\begin{center}
	\begin{tabular}{lccccc}\hline
			\mdseries{Alloy} & \mdseries{Co}  & \mdseries{Cr} &
			\mdseries{Fe} & \mdseries{Mn} & \mdseries{Ref.}\\
			\hline
			Co$_{25}$CrFeMnNi$_{15}$ & $1.2 \pm 0.1$ & $3.9 \pm 0.1$ & $2.4 \pm 0.1$ &
			$8.2 \pm 0.2$ & present work
			\\
			Co$_{15}$CrFeMnNi$_{25}$ & $2.1 \pm 0.1$ & $5.6 \pm 0.1$ & $4.2 \pm 0.1$ &
			$13.3 \pm 0.6$ & present work
			\\
			CoCrFeMnNi               & $1.9 \pm 0.2$ & $5.0 \pm 0.6$ & $3.4 \pm 0.5$ &
			$9.0 \pm 1.0$ & \cite{Gaertner2018}\\
			\hline
		\end{tabular}
	\end{center}
	\label{tab:OutDiffusion}
\end{table}

\noindent
In both alloys, Mn is found to be the fastest element and Co the slowest one 
(note that Ni tracer diffusion was not measured in the present work).
Table \ref{tab:OutDiffusion} suggests further that the diffusion rates of all
investigated elements are increased by up to $75\,\%$ after alloying the
opposite amount of Co and Ni, while keeping an equiatomic ratio of the other three elements. A
direct comparison of the tracer and chemical profiles in the vicinity of the
Matano plane reveals immediately significantly different scales on which volume
diffusion could reliably be followed. Indeed, if $^{54}$Mn tracer diffusion is
measurable in the region of $\pm400\,\mu$m and it is $\pm100\,\mu$m for $^{57}$Co, the
chemical changes are confined within $\pm80\,\mu$m from the initial interface,
see Fig. \ref{fig:Diffusion}.

\subsubsection{Combined tracer-interdiffusion measurements}

\begin{figure}[ht]
  \includegraphics[width=.9\linewidth]{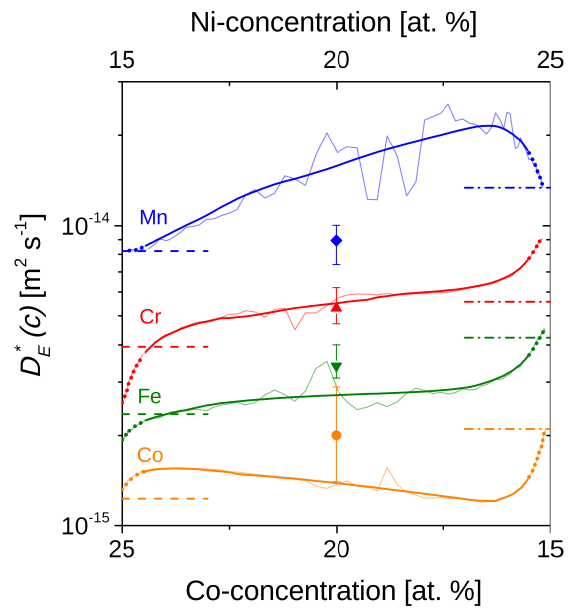}

\caption{\footnotesize Tracer diffusion coefficients
   of Co, Cr, Fe and Mn as a function of the Co-concentration
   (represented by solid lines) compared with the tracer diffusion
   coefficients measured in the sides with constant constituent concentration
   (represented by dashed lines for the Co-rich side and dashed-dot lines for
   the Ni-rich side) and with the tracer diffusion coefficients measured in single
   crystalline CoCrFeMnNi HEA \cite{Gaertner2018} (represented by filled
   symbols). The dotted lines show the determined tracer diffusion coefficients
   for the near end-member compositions.}
  \label{fig:Belova-Murch}
\end{figure}

In order to analyze the tracer concentration profiles of all constituting
elements developed at the original interface of the pseudo-binary couple, the
Gaussian solution, Eq.~(\ref{eq:Gauss}), is invalid due to a strong
chemical driving force. The concentration-dependent tracer diffusion
coefficients can be determined using the framework of a \textit{thin layer isotope sandwich
configuration} \cite{Belova2015},

\begin{equation}
D^*_E(c) =
- \frac{\frac{(y+a)}{2t}-\frac{G_E(y)}{c_E(y)}}{\frac{\partial \ln
		c^*_E(y)}{\partial y}-\frac{\partial \ln c_E(y)}{\partial y}} 
\label{eq:BelovaMurch}
\end{equation}

\noindent where $c_E(y)$ is the concentration of the element $E$, $c^*_E(y)$ the
tracer concentration of the same element, $G_E(y)$ is the value proportional to
the flux of the element $E$. The value of the parameter $a$ corresponds to a
misfit of the $y$ scales for the chemical, $c_E(y)$, and the tracer, $c^*_E(y)$,
profiles and $a$ can be determined from the condition that the tracer diffusion
coefficient $D^*_E(c)$ is positive at all $y$. Making use of the Sauer-Freise
method \cite{Sauer1962} the factor $G_E(y)$ can be determined \cite{Belova2015},

\begin{eqnarray}
G_E(y) & = & \frac{c_{\text{R}}-c_{\text{L}}}{2t}\times \left[(1-Y_E)\int_{-\infty}^{y}Y_Edy+ \right. \nonumber \\ 
& &  \left. Y_E\int_{y}^{\infty}(1-Y_E)dy\right]
\label{eq:flux}
\end{eqnarray}

\noindent where $Y_E(y)=(c_E(y)-c_{\text{L}})/(c_{\text{R}}-c_{\text{L}})$ is
the reduced concentration of the given chemical element $E$. A variation of the
molar volume with composition is neglected, which is a reasonable approximation
in the present case.\\
\noindent
In Fig. \ref{fig:Belova-Murch}, the determined concentration-dependent
tracer diffusion coefficients are shown as straight lines. The
profiles show peak-like artefacts resulting from the amount of data points and
the fitting of the parameter $a$ in Eq. \ref{eq:flux}. Since the interdiffusion
coefficients determined by the Sauer-Freise method are typically prone to large
uncertainties for the compositions close to the end-members, the corresponding
values are indicated as dotted lines in Fig. \ref{fig:Belova-Murch}.

\noindent
The concentration dependent tracer diffusion coefficient of Co shows a S-shaped
trend and the independently measured diffusion coefficients for the end-member
 concentrations, represented by the dashed and dashed-dotted lines for the Co-rich and Ni-rich sides, respectively, are in a good agreement
with the determined trends within the typical accuracy of about $20\,\%$ for the
tracer experiments.\\
\noindent
The present analysis predicts a clear decrease of the Co tracer coefficient
with decreasing Co concentration from $23$ to $17\,\text{at.}~\%$. Contrary to
that, the other elements show an increase of the diffusion rate in this concentration range, especially if the near-end member concentrations are not
included in the analysis. This Co behavior is the subject of an ongoing work in a pseudo-binary couple with Co$_{23}$CrFeMnNi$_{17}$ and
Co$_{17}$CrFeMnNi$_{23}$ HEAs.\\
\noindent
The independently determined tracer diffusion coefficients, i.e. those for the
end-members, Eq.~\ref{eq:Gauss}, and along the diffusion path, Eqs.
\ref{eq:BelovaMurch} \& \ref{eq:flux}, for Mn are in a good agreement, while the
deviations are larger for Cr and Fe.\\
\noindent
The tracer diffusion coefficients measured in
equiatomic single crystalline CoCrFeMnNi HEA \cite{Gaertner2018} are shown as
filled symbols in Fig. \ref{fig:Belova-Murch}. In case of Co and Fe
the volume diffusion coefficients are higher than the corresponding coefficients
$D^*_E(c)$, however, the values overlap accounting for the
measurement uncertainties. The coefficient $D^*_{\text{Cr}}
(c)$ is in very good agreement with the single crystal
data for Cr, while the Mn volume diffusion coefficient in the equiatomic state
is somewhat overestimated. This relatively large deviation of about $50\,\%$
may probably result from the strong up-hill diffusion contribution.

\section{Diffusion simulations}
\subsection{Model description}
\subsubsection{Pair-wise diffusion model}  \label{Pair-wise diffusion model}
In the present work, a generalized diffusion model, which is independent of
a reference element, is proposed. The fluxes given in the lattice-fixed frame
are transformed into the laboratory-fixed frame following the approach of
Boettinger et al.~\cite{Boettinger2016} who derived the flux equations for the
binary case. Extending it to the multi-component case and using the assumption
that all partial molar volumes are equal and independent of composition, as it
was done for the analysis of the experimental data above, the fluxes can be
written in the following pair-wise form (for a short derivation see the Appendix
A, a detailed derivation and analysis of the resulting diffusion equation will
be published elsewhere):

\begin{eqnarray}
\tilde{J}_i & = & -\frac{1}{2}\sum_{j}^{n}x_ix_jM_{ij}(\nabla \mu_i- \nabla \mu_j) = \nonumber \\ 
& & -\frac{1}{2}\sum_{j}^{n}x_ix_jM_{ij}\nabla\tilde{\mu}_{ij}
\end{eqnarray}
\noindent with $x_i$ as the mole fraction of element $i$, $M_{ij}$ is the
concentration-dependent pair-exchange mobility and $\mu_i$ the chemical
potential of element $i$. The change of composition is then given as:

\begin{equation}
\frac{\partial x_i}{\partial t}=-\nabla \tilde{J}_i=\frac{1}{2}\nabla
\sum_{\substack{j=1\\j\neq i}}^{n}x_ix_jM_{ij}\nabla\tilde{\mu}_{ij}.
\label{Pair-wise diffusion equation}
\end{equation}

\noindent The sum is taken over all pairs of elements. The key point of the
present ansatz is that the thermodynamic driving force is given by the gradient
of the \emph{difference} of the chemical potentials of each pair:

\begin{equation}
\tilde{\mu}_{ij}=\mu_i-\mu_j=\frac{\partial G}{\partial x_i} -\frac{\partial G}{\partial x_j}. 
\end{equation} 

\noindent $M_{ij}$ is the pair-exchange mobility of element $i$ and $j$. It
represents the exchange rate of solutes through a unit area within the reference
volumina at the continuum scale, and should not be confused with an atomistic pair-exchange mechanism. It may be due to an atomistically defined vacancy mechanism with 'many' individual jumps, or other mechanisms. The key idea is to decompose a general multi-solutal diffusion process in pairs of exchange processes with a common reference in the diffusion potential. \\
\\
\textbf{Pair-exchange mobility $M_{ij}$}\\
The pair-exchange mobility can be derived by transforming the intrinsic fluxes
in the lattice-fixed frame of reference $J_i = -M_i\frac{x_i}{V_m}\nabla \mu_A$
into the laboratory frame $\tilde{J_i}$ including the velocity with which the
frames move with respect to each other. Rewriting the resulting flux as
pair-wise contributions, given in Eq.~\ref{Pair-wise diffusion equation}, the
pair-exchange mobility is defined as:
\begin{equation}
M_{ij}=x_iM_j+x_jM_i+\sum_{\substack{k=1\\k\neq i\\k\neq j}}^{n}x_k(M_i+M_j-M_k)
\label{Atomic_mobility_Ansatz}
\end{equation}
with $M_k$ as the concentration-dependent atomic mobility of element $k$. Note
that for large differences in the mobilities of different elements, in
particular $M_k$~$\gg$~$M_i$,~$M_j$, the pair mobility $M_{ij}$ can become
negative. In general, this is not a problem, since one has to ensure that the
diffusion matrix is positively defined for consistency with the second law of
thermodynamics. See also recent discussion in~\cite{Chen2018}. \\
It can be directly seen that the introduced pair-exchange mobilities are
symmetric:
$M_{ij}$ = $M_{ji}$. In the binary case the pair-exchange mobility reduces to
$M_{ij}=x_iM_j+x_jM_i$. In higher order systems, additional to the binary
pair-exchange mobility, a term over all other elements except $i$ and $j$
(second part of Eq.~\ref{Atomic_mobility_Ansatz}) influences the pair-exchange
mobility between $i$ and $j$. The diagonal terms ($M_{ii}$) are not defined. It
is shown in Appendix B, that the generalized pair-wise diffusion model reduces
to the DICTRA model~\cite{Andersson1992} in the dilute solution limit.\\
\\
\textbf{Atomic mobility $M_i$}\\
The pair-exchange mobility $M_{ij}$ can be constructed from the composition
dependent atomic mobilities $M_i$, see Eq.~\ref{Atomic_mobility_Ansatz}.
In 1992 Andersson and {\AA}gren~\cite{Andersson1992} proposed to store the
temperature and composition dependent atomic mobilities in CALPHAD-type kinetic
databases and model the temperature dependence as~\cite{Andersson1992,Campbell2001}:
\begin{equation}
M_i=\frac{1}{RT}M_i^0\exp{\bigg(-\frac{Q_i}{RT}\bigg)}\left[^{\text{mag}}\Gamma_i\right].
\end{equation}
R is the gas constant, T the temperature, $M_i^0$ the frequency factor, $Q_i$
the activation energy for diffusion and $^{\text{mag}}\Gamma_i$ the magnetic
contribution (set to unity in the present case). It is customary in most kinetic
databases to include the composition dependence in $Q_i$ using Redlich-Kister
polynomials, while $M_i^0$ is equal 1~\cite{Campbell2001,
Redlich1948}\footnote{$Q_i$ will be fitted using $Q_i=a_i+b_i\cdot(T-T_0)$,
while $T_0$ corresponds to the reference temperature and $T$ corresponds to the
measuring temperature. Here, the reference temperature is equal to the
measuring temperature $T=T_0$, so the fitting parameter $b_i$ is undefined and
the fitting parameter $a_i$ is given.}:
\begin{equation}
Q_i = \sum_jx_jQ_i^j+ \sum_p\sum_{j>p}x_px_j\sum_kA_i^{pj}(x_p-x_j)^k
\end{equation}
with $Q_i^j$ and $A_i^{pj}$ as fit parameters. Nearly every temperature and
composition-range in most phases is covered either by assessments or by
extrapolating the existing data to the given system using the described scheme
for composition and temperature dependence.\\
In this paper not only kinetic databases are used to describe the atomic
mobilities but also direct use is made of the experimentally measured tracer
diffusion coefficients $D_i^*$, applying the Einstein relation:
\begin{equation}
M_i(c)=\frac{D_i^*(c)}{RT}.
\label{Einstein relation}
\end{equation}
Taking advantage of the particular set-up of the modified tracer-interdiffusion
couple (MTIC) experiment, i.e. the diffusion measurements within the
interdiffusion zone and in the unaffected end-members (see Fig.~\ref{fig:Diffusion}), two
different data repositories, applicable for the diffusion simulations in the
given composition-range, are established: \\
1.~The tracer diffusion coefficients determined from the measurements in the
unaffected end-members (see Table~\ref{tab:OutDiffusion}) are linearly
interpolated and stored in the data repository called MTIC-Lin (the functions
are given in Appendix C). \\
2.~The concentration-dependent tracer diffusion coefficients determined along
the interdiffusion path using the Belova-Murch approach \cite{Belova2015} (shown
in Fig.~\ref{fig:Belova-Murch}) which cover the investigated composition range and make an interpolation redundant, are
directly used.
This data repository is referred to as MTIC-BM in the following.\\
These two approaches allow a direct use of the experimentally measured kinetic
data in the diffusion simulations. It is possible to rewrite the data in
Redlich-Kister polynomials as it is done for CALPHAD-type kinetic databases (in Appendix
C the Redlich-Kister polynomials are given as an example for MTIC-Lin).\\

\subsubsection{Tracer diffusion simulations}
\textbf{Thermodynamic and kinetic model for tracer atoms}\\
In the following simulations the mass effect of isotopes on diffusion is
neglected. Therefore the radioactive tracer atoms are chemically
indistinguishable from the stable isotopes (non-tracer) of the same species and their thermodynamic
and kinetic properties are considered to be the same. The total composition of one species is then given by:
\begin{equation}
x_{i,\text{tot}}=x^*_{i}+x_{i}.
\end{equation}
$x^*_{i}$ is the tracer concentration of species $i$ and $x_{i}$ is the amount
of non-tracer atoms of species $i$.
Thermodynamic and kinetic properties are always evaluated with respect to
$x_{i,\text{tot}}$. Pair-wise diffusion is applied to $x_{i,\text{tot}}$ and
does not distinguish between $x^*_{i}$ and $x_{i}$. Therefore the pair-wise model does
not take into account self-diffusion of tracer atoms (the corresponding
difference of the chemical potential gradients is $\mu_{i^*}-\mu_i = 0$). \\
\\
\textbf{Self-diffusion model}\\
To take into account self-diffusion of tracer atoms, a second diffusion
model is used. There is no explicit exchange between different isotopes of the same element within the pair-exchange model as mentioned above. With the aim to reproduce experimentally measured tracer profiles
one can safely assume that the amount of tracer atoms used in the experiments is
too small to influence the overall concentration in a measurable amount (for
EPMA analysis), see Eq.~\ref{eq:self_diffusion} that was verified by direct
estimates of the absolute concentrations of tracer atoms in a diffusion experiment~\cite{Divinski2001}. The key point is that self-diffusion is measured and no
impurities - which may effect the vacancy concentration especially in
non-metallic systems - are introduced by application of tracer solutions.
Therefore cross terms can be neglected and one can assume an ideal solution model. Self-diffusion can then be described
with Fick's second law:
\begin{equation}
\frac{\partial x_i^*}{\partial t}=\nabla \left[ D_i \nabla x_i^* \right].
\label{eq:self_diffusion}
\end{equation}
$x_i^*$ is the concentration of the tracer atoms of species $i$ and $D_i$ is the
self-diffusion coefficient of element $i$ (composition dependence is evaluated
with respect to the total composition). Applying the fluctuation-dissipation
theorem \cite{Kubo1966} $D_i$ shall be identified with $D_i^*$ in
Eq.~\ref{Einstein relation}.

\begin{figure*}
	\centering
	\includegraphics[width=0.9\textwidth]{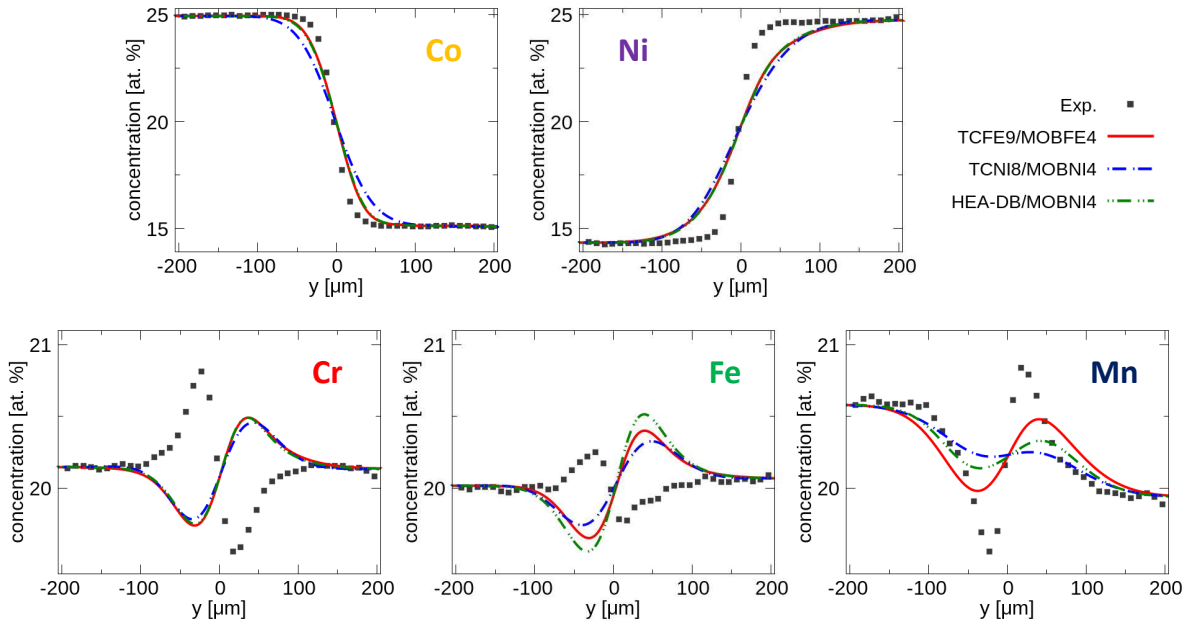}
	\caption{\footnotesize Comparison between experimentally measured
		interdiffusion profiles and simulated ones using the three different
		thermodynamic/kinetic databases (compare Table~\ref{tab:Databases}) for all
		five elements (Co, Cr, Fe, Mn and Ni) after 48~h at 1373~K. }
	\label{fig:Interdiffusion_Sim_profiles}
\end{figure*}

\subsection{Simulation results}
In the following simulations the annealing time and duration were
chosen for comparability as in the experiment (1373~K, 48~h). The pair-wise ansatz for
chemical diffusion and tracer diffusion in varying chemical composition are
solved explicitly using adaptive time stepping with a constant grid spacing ($1
\, \mu m$). The box size of the 1D simulations were chosen to represent a
semi-infinite sample where the concentrations at the ends are not affected by
interdiffusion. Fixed concentrations were taken as boundary conditions.
The initial amount of tracer ($x_{i}^*$) does not influence the profiles as long
as it is smaller than the overall concentration of the given element ($x_{i}
^*\leq x_{i,\text{tot}}$). To minimize numerical errors, the initial tracer
distribution amount was chosen as $x_{i}^* = 0.25\cdot\delta(y)$ for Co, Cr, Fe
and Mn, with $\delta(y)$ as the Dirak-delta function and the distance $y$.\\
In the sublattice representation of CoCrFeMnNi all elements in the fcc lattice are
on the substitutional sublattice and the interstitial sublattice is only occupied by
vacancies (Co, Cr, Fe, Mn, Ni)$_1$(Va)$_1$. Three different sets of databases
(thermodynamic + kinetic database) are investigated in the following simulations
(summarized in Table~\ref{tab:Databases}). TCNI8/MOBNI4~\cite{TCNI_dat} and
TCFE9/MOBFE4~\cite{TCFE_dat} were developed for alloys with Ni respectively Fe
as principal element. Another thermodynamic database was developed by
Hallstedt's group (in the following abbreviated by HEA-DB) especially for HEAs~\cite{Haase2017}. Due to the lack of an explicit HEA kinetic database it was
also combined with the MOBNI4~\cite{TCNI_dat} database. The HEA databases
developed by Thermo-Calc~\cite{TCHEA_dat} were not considered in this work, due
to the availability of the HEA-DB.

\begin{table}[h]
	\footnotesize
	\caption{\footnotesize Combinations of thermodynamic and kinetic databases used in the following simulations. For the combination HEA-DB/MOBNI4, Ni as reference element is given in brackets, because only the kinetic database is based on a reference element: Ni.}
	\begin{center}
		\begin{tabular}{cccc}
			\hline
			\mdseries{Thermodynamic Database} & \mdseries{Kinetic Database} & \mdseries{Reference Element} \\
			\hline
			TCNI8~\cite{TCNI_dat} & MOBNI4~\cite{TCNI_dat}  & Ni \\
			TCFE9~\cite{TCFE_dat} & MOBFE4~\cite{TCFE_dat}  & Fe \\
			HEA-DB (Hallstedt's group)~\cite{Haase2017}& MOBNI4~\cite{TCNI_dat} & (Ni) \\
			\hline
		\end{tabular}
	\end{center}
	\label{tab:Databases}
\end{table}

\subsubsection{Interdiffusion}
The experimentally measured and simulated interdiffusion profiles using the
three different database combinations given in Table~\ref{tab:Databases} are
shown in Fig.~\ref{fig:Interdiffusion_Sim_profiles}. \\
\\
\textbf{Co and Ni}\\
The simulated profiles for Co and Ni are for all database combinations
considerably shallower than the experimentally measured profiles. Using
TCNI8/MOBNI4 and HEA-DB/MOBNI4 the predicted profiles nearly coincide for both
elements.
For Co the simulated profile using TCFE9/MOBFE4 is closer to the experimental
result than the others. For Ni, all database combinations provide nearly similar
descriptions strongly deviating from the experimental profile.\\

\begin{figure*}
	\centering
	\includegraphics[width=0.9\textwidth]{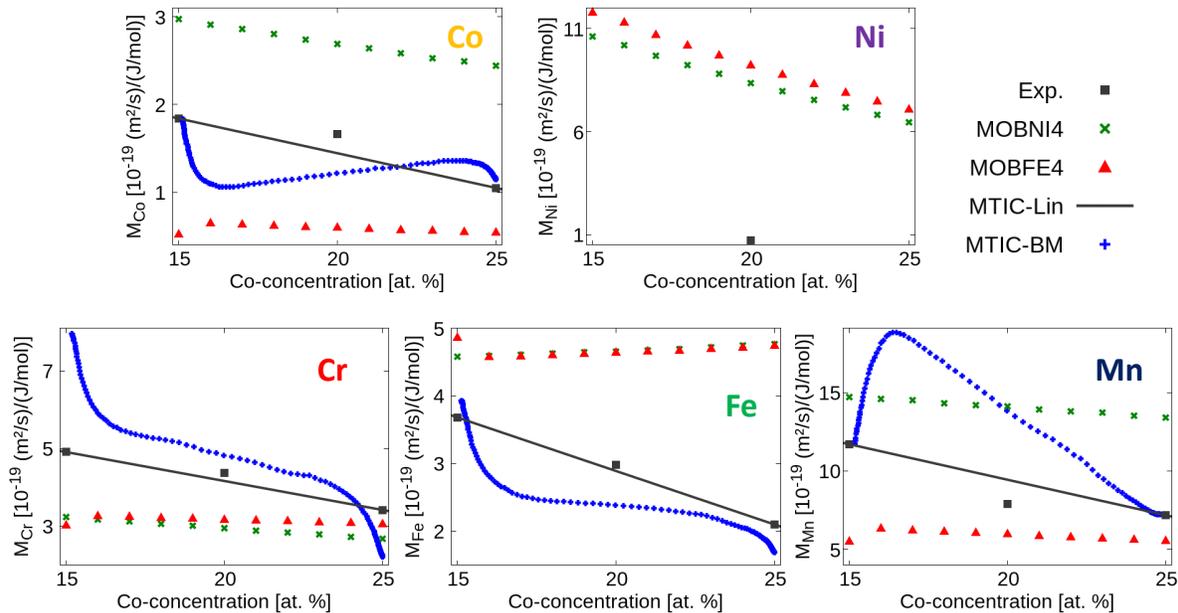}
	\caption{\footnotesize Composition dependent atomic mobilities for Co, Cr, Fe, Mn and Ni. Cr, Fe and Mn concentration is assumed to be constant ($20$~at.$\,\%$
		each). The grey squares are the values determined from
		the experiments in constant composition (compare Table~\ref{tab:OutDiffusion})
		and a linear interpolation between the end points is shown as solid line (MTIC-Lin). MTIC-BM (blue crosses) represents the result from the modified
		tracer-interdiffusion couple.}
	\label{fig:Comparison atomic mobilities}
\end{figure*}

\textbf{Cr, Fe and Mn}\\
For the three elements Cr, Fe and Mn, uphill diffusion was observed in the
experiment and is also observed in the simulations. For Cr and Fe the simulated
results reveal uphill diffusion in the opposite direction than the experiment
(Experiment: uphill diffusion along the concentration gradient of Ni;
Simulations: uphill diffusion against the concentration gradient of Ni - see
Fig.~\ref{fig:Interdiffusion_Sim_profiles}). The different databases are only
distinct from each other in the magnitude of the uphill profile. For Mn the
simulated profiles using TCFE9/MOBFE4 and HEA-DB/MOBNI4 show uphill diffusion in
the same direction as the experiment while for TCNI8/MOBNI4 the profile is rather flat.\\
\\
It should be noted that the database combinations TCNI8/MOBNI4 and TCFE9/MOBFE4
were not developed for the equiatomic composition range. But although the HEA-DB
thermodynamic database was developed for the given composition range,
simulations using this database cannot reproduce the experimentally measured
profile, which might be due to the not fitting kinetic database.

\begin{figure*}
	\centering
	\includegraphics[width=0.9\textwidth]{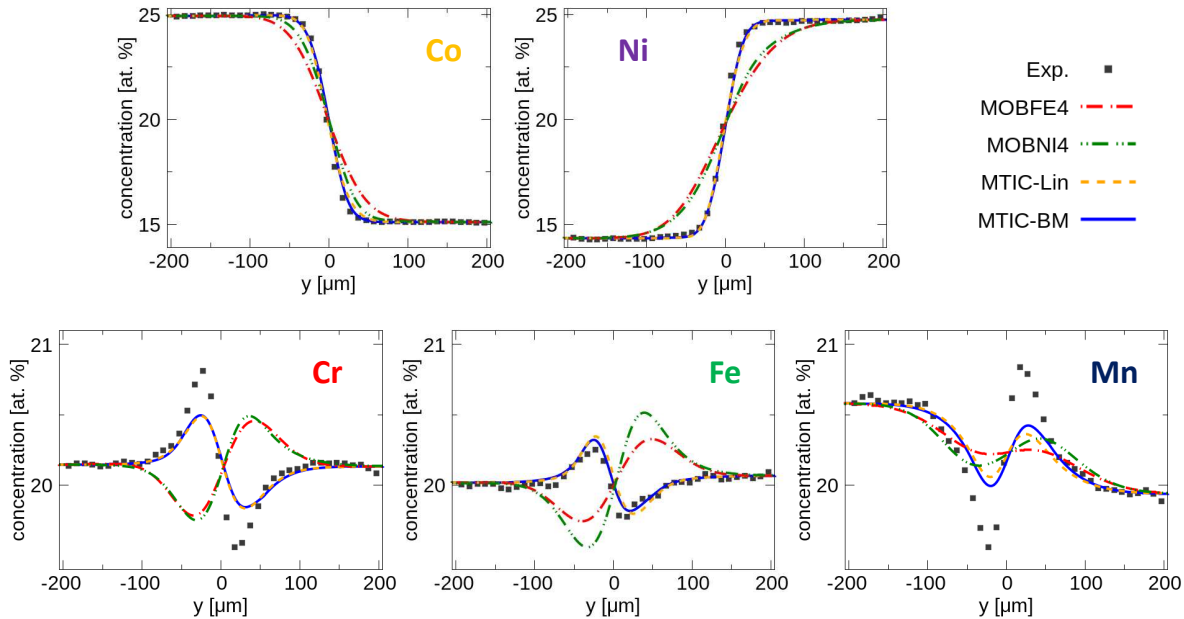}
	\caption{\footnotesize Comparison between experimentally obtained
		interdiffusion profiles and simulated ones using the HEA-DB~\cite{Haase2017}
		combined with different atomic mobility databases (compare
		Fig.~\ref{fig:Comparison atomic mobilities}) for all five elements (Co, Cr, Fe,
		Mn and Ni) after 48~h at 1373~K.}
	\label{fig:Switch_uphill}
\end{figure*}

\subsubsection{Comparison of atomic mobility databases to experimental results}

Atomic mobilities which are directly related to the self-diffusion coefficients
by the Einstein relation ($D_i=RTM_i$) and their composition dependence play a
significant role in interdiffusion, see Eq.~\ref{Pair-wise diffusion
equation}, and in tracer diffusion, see Eq.~\ref{eq:self_diffusion}. In
Fig.~\ref{fig:Comparison atomic mobilities} the composition dependence of the
atomic mobilities from different sources is presented with respect to the
variation of the Co/Ni concentration. Cr, Fe and Mn concentrations are assumed
to be constant ($20$~at.$\,\%$) while the Co concentration is increased on the
expense of Ni. This represents the simplified version of the zone with
neglection of uphill diffusion.\\
The atomic mobilities determined using the modified tracer-interdiffusion couple
(MTIC) along the interdiffusion path and shown in Fig.~\ref{fig:Belova-Murch}
are represented by blue crosses. To convert it into a function only depending on
the Co/Ni concentration, the result was averaged over the Co concentration (assuming
the deviations of Cr, Fe and Mn from $20$~at.$\,\%$ are negligible).
The tracer volume diffusion coefficients listed in Table~\ref{tab:OutDiffusion}
for Co$_{25}$CrFeMnNi$_{15}$ and Co$_{15}$CrFeMnNi$_{25}$ are shown by grey
squares. To obtain a continuous composition dependence, they were linearly
interpolated along the diffusion path (grey solid line).\\
For Co the atomic mobilities are highest obtained from MOBNI4 database and they
are decreasing with increasing Co concentration. Atomic mobilities from MOBFE4
database are the lowest ones and those ones determined from the experiments are
in between. For Cr the atomic mobilities from MOBFE4 and MOBNI4 are similar to
each other and smaller than those ones obtained from the experiments. The same
accounts for the atomic mobility of Fe, but in this case the databases offer
higher values than those ones determined from the experiments. In case of Mn
the atomic mobilities from MOBNI4 are higher than those ones from MOBFE4. Again
the experimentally determined atomic mobilities are intermediate.\\
Ni tracer diffusion was not measured in the combined tracer and interdiffusion
experiment, but there are data for the equiatomic alloy (only tracer) at the
same temperature. Since the composition dependence of the Ni tracer diffusion
coefficient was not evaluated, it is taken in the following simulations as a
constant. A comparison of the measured Ni tracer diffusion coefficient in the
equiatomic alloy to the data from the MOBNI4 and MOBFE4 database is also shown
in Table~\ref{tab:Ni_self_diffusion_coefficients}. In both databases Ni
diffusion is predicted as one order of magnitude faster than it is measured in
the equiatomic alloy.

\begin{table}[h]
	\footnotesize
	\caption{\footnotesize Comparison of Ni self-diffusion coefficients in equiatomic CoCrFeMnNi at 1373~K.}
	\begin{center}
		\begin{tabular}{cc}
			\hline
			Source & D* ($10^{-15}$~m$^2$s$^{-1}$)\\
			\hline
			MOBNI4 & 9.54   \\
			MOBFE4 & 10.50\\
			Experiment& 0.83\\
			\hline
		\end{tabular}
	\end{center}
	\label{tab:Ni_self_diffusion_coefficients}
\end{table}	 

\subsubsection{Influence of the atomic mobilities on the interdiffusion profiles}

For the following simulations only HEA-DB~\cite{Haase2017} is considered for
thermodynamics and combined with the different CALPHAD-type kinetic databases
(MOBFE4 and MOBNI4) and the new kinetic databases (MTIC-Lin and MTIC-BM)
determined from experiments. For MTIC-Lin and MTIC-BM the Ni diffusivity is
taken as constant ($D^*_{\text{Ni}} =$ 8.34$\times$10$^{-16}$ m$^2$s$^{-1}$)
determined from the tracer diffusion experiment in the equiatomic sample.
The resulting interdiffusion profiles are shown in Fig.~\ref{fig:Switch_uphill}.\\
\\
\noindent\textbf{Co}\\
In case of Co the MOBNI4 database provides the highest atomic mobilities while
the MOBFE4 database gave the smallest ones. The atomic mobilities obtained from the
experiments were in between (compare Fig.~\ref{fig:Comparison atomic
mobilities}). The simulated diffusion profiles do not represent this order. Both
simulations using the atomic mobilities from the databases were significantly
faster than the experiment. Using the atomic mobilities determined with the MTIC
approach (MTIC-Lin and MTIC-BM data repository), both reproduce the experimental
result very well. This highlights that the resulting Co interdiffusion profile
is significantly influenced by the mobilities of the other elements. We
highlight that exactly such cross-correlations are inherent for the new ansatz
proposed in the present paper, Eq.~\ref{Atomic_mobility_Ansatz}.\\
\\
\textbf{Ni}\\
The largest deviations between the atomic mobilities from the databases and the
experiment were found for Ni, see Table~\ref{tab:Ni_self_diffusion_coefficients} and Fig.~\ref{fig:Comparison atomic
mobilities}, which is reflected in the final Ni interdiffusion profile. Using
the kinetics from the databases (MOBFE4 and MOBNI4) results in significantly flatter profiles than
the experimentally measured one, while using the Ni self-diffusion coefficient
as a constant, obtained from the tracer experiment in the equiatomic alloy, reproduces it very well, Fig.~\ref{fig:Comparison atomic
mobilities}. \\
\\
\textbf{Cr, Fe and Mn}\\
Applying the MTIC-Lin, as well as the MTIC-BM kinetic data, uphill diffusion for Cr
and Fe reverts compared to the profiles obtained using the CALPHAD-type databases. For
Cr the measured diffusivities (MTIC-Lin and MTIC-BM) are slightly higher than those
ones from the databases, and for Fe it is the other way around. Because this
inversion can only be seen when Ni is slowed down, the cross effects with Ni
play a key role for diffusion of Cr and Fe.\\
For Mn the uphill diffusion is in the same direction as in the experiment
although it is not as distinct as in the experiment. Using the MTIC-BM approach
gives a slightly better result than using MTIC-Lin.

\begin{figure*}
	\centering
	\includegraphics[width=0.9\textwidth]{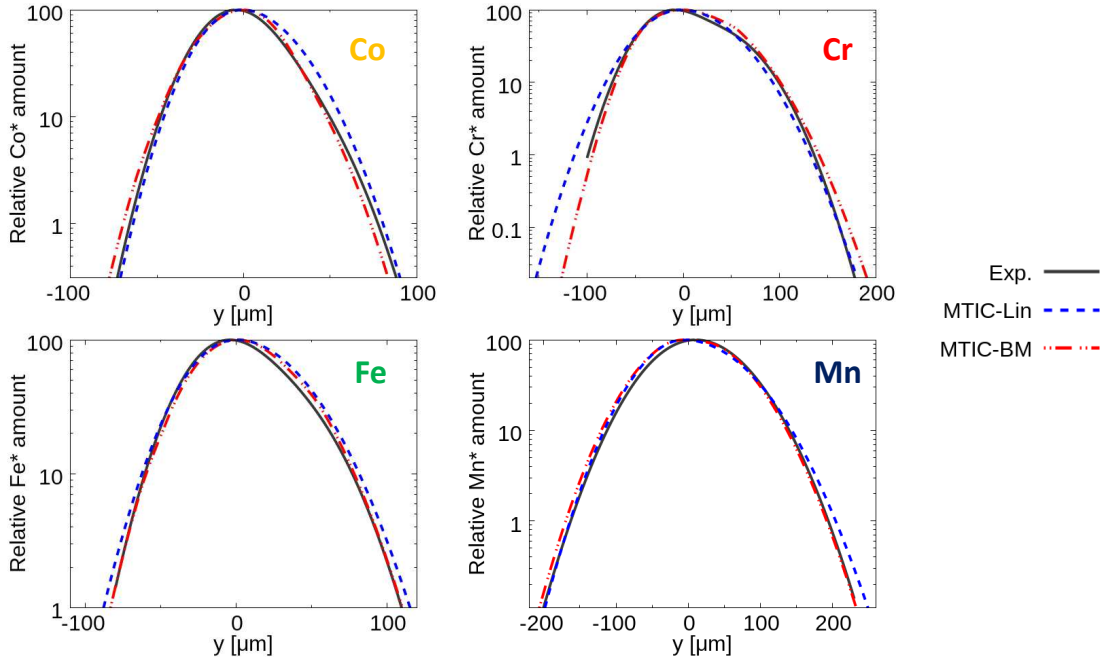}
	\caption{\footnotesize Comparison between experimentally obtained tracer
		profiles and simulated ones in the interdiffusion zone. The thermodynamic
		database HEA-DB was used and combined with two different composition dependent
		atomic mobility data repositories: MTIC-Lin and MTIC-BM (compare section~\ref{Pair-wise diffusion model} - Atomic mobility $M_i$). }
	\label{fig:Tracer final}
\end{figure*}

\subsubsection{Tracer diffusion simulations}
It was shown in the previous section that the measured self-diffusion
coefficients in this work play a key role for the interdiffusion profiles. The
advantage of the combined tracer-interdiffusion experiment is not only the
determination of the composition dependent self-diffusion coefficients but also
the measured tracer profiles that can be compared to the simulated ones. \\
The experimental and simulated tracer profiles in the interdiffusion zone are
shown in Fig.~\ref{fig:Tracer final}. The results are given for the
thermodynamic database HEA-DB combined with the atomic mobilities from MTIC-Lin
and MTIC-BM. For all four elements the differences between the curves are small.
For Co and Fe the whole profile is better represented using the MTIC-BM data
repository. For Cr the Ni rich (right) side is better reproduced using MTIC-Lin,
while the Co rich (left) side is better with MTIC-BM. For Mn the Co rich (left)
side is slightly better using MTIC-Lin while the Ni rich (right) side is
perfectly represented using MTIC-BM.

\section{Summary}

In the present work, the concentration dependent tracer diffusion coefficients
and the tracer diffusion coefficients of the unaffected end-members were
determined for the first time using a modified tracer-interdiffusion couple
approach. All the unaffected end-member tracer diffusion coefficients increase
up to $75\,\%$ with decreasing Co-concentration along the diffusion path. The
concentration dependent tracer diffusion coefficient of Co was found to
follow a non-monotonous behaviour, which is influenced by the up-hill diffusion
of Mn.
The end-member diffusion coefficients of Co, Fe and Mn are in a good agreement
with the determined trends for intermediate concentrations, while Cr shows
larger deviations.\\
Based on the experimentally determined atomic mobilities, the prediction of the
experimental results by diffusion simulations using a new ansatz for the
generalized diffusion model is more exact than using other existing kinetic databases. 
Using the experimental results of the modified tracer-interdiffusion couple
(MTIC) as the mobility database and the thermodynamic database by
Bengt-Hallstedt (HEA-DB) - which was developed for the given near-equiatomic compositions - 
the simulations predict the interdiffusion-profiles as well as the
tracer-profiles very well. For the elements without an initial
concentration gradient the direction of the up-hill diffusion agrees perfectly
with the experiment when the MTIC-Lin or MTIC-BM results are used, while
it is reverse if the existing databases with Fe or Ni as the
reference element are applied.\\
Accounting for significantly different scales on which tracer and chemical
diffusion near the Matano plane could be followed, Fig.~\ref{fig:Diffusion}, we
conclude that the concept of 'sluggish' diffusion - if at all - may be
applicable for the description of chemical instabilities in HEAs, but
definitely not for tracer diffusion in these alloys.

\section*{Acknowledgement}

We would like to thank the Deutsche Forschungsgemeinschaft
(DFG), within the research projects DI 1419/13-1 \& STE 116/30-1), for the
financial support.

\section*{Appendix A - Derivation of the pair-wise diffusion model}
\renewcommand{\theequation}{A\arabic{equation}}
\setcounter{equation}{0}

In a multi-component system the following relations are given and used in this
derivation: $x_j = c_jV_m$, $ V_m = \sum_{j=1}^{n}x_j\tilde{V}_j$,
$\sum_{j=1}^{n}\tilde{V}_jc_j = 1$, $\sum_{j=1}^{n}\tilde{V}_jdc_j = 0$ and the
Gibbs-Duhem relation ($V_m$ is the molare volume and $\tilde{V}_j$ the partial
molare volume with respect to element $j$).\\
Starting from the mass conservation equations as in \cite{Boettinger2016}:
\begin{align}
\frac{\partial c_j}{\partial t}+\nabla J_j + \nabla (c_j*v)&=0\\
\end{align}
and replacing the velocity by:
\begin{align}
\nabla v=-\sum_{i=1}^{n}\nabla J_jV_j
\end{align}
results in:
\begin{align}
\frac{\partial c_j}{\partial t}=-\nabla \sum_{\substack{i=1 \\ i \neq j}}^nV_ic_iJ_j+ \nabla(c_j\sum_{\substack{i=1\\i\neq j}}^{n}J_iV_i)
\label{c_j_change}
\end{align}
Inserting the intrinsic flux $J_i=-\widehat{M}_ic_i\frac{\partial \mu_i}{\partial_z}$ and
making use of the Gibbs-Duhem equation:
\begin{align}
\frac{\partial c_j}{\partial t}=-\nabla\sum_{\substack{i=1 \\ i \neq j}}^n\bigg[\sum_{\substack{k=1\\k \neq j}}^{n}c_ic_k\widehat{M}_jV_k\nabla\mu_i+c_ic_j\widehat{M}_iV_i\nabla\mu_i\bigg]
\end{align}
Multiply by $1=\sum_{i=1}^{n}x_i=V_m\sum_{i=1}^{n}c_i$:
\begin{eqnarray}
\frac{1}{V_m}\frac{\partial c_j}{\partial t} &=&-\nabla\sum_{\substack{i=1 \\ i \neq j}}^n\bigg[\sum_{\substack{k=1\\k \neq j}}^{n}c_ic_k\widehat{M}_jV_k\nabla\mu_i+c_ic_j\widehat{M}_iV_i\nabla\mu_i\bigg] \times \nonumber \\
& & \sum_{l=1}^{n}c_l
\end{eqnarray}
Finally rewrite this term into pair-interactions:
\begin{eqnarray}
\frac{\partial x_j}{\partial t}&=&\frac{1}{V_m}\nabla\sum_{\substack{i=1 \\ i \neq j}}^n\bigg[\bigg(\sum_{\substack{k=1\\k\neq j\\k\neq i}}^{n}x_ix_jx_k(\widehat{M}_iV_i-\widehat{M}_kV_k+\widehat{M}_jV_k)  \nonumber \\
&+& x_ix_j(x_i\widehat{M}_jV_i+x_j\widehat{M}_iV_i)\bigg)\nabla(\mu_j-\mu_i)\bigg]
\end{eqnarray}
Assume $V_i=V_j=V_k=V_m$:
\begin{eqnarray}
\frac{\partial x_j}{\partial t} &=& \nabla \sum_{\substack{i=1 \\ i \neq j}}^n\bigg[x_ix_j\bigg(x_i\widehat{M}_j+x_j\widehat{M}_i+ \nonumber \\
&&\sum_{\substack{k=1\\k\neq j\\k\neq i}}^{n}x_k(\widehat{M}_i+\widehat{M}_j-\widehat{M}_k)\bigg)\nabla(\mu_j-\mu_i)\bigg]
\end{eqnarray}
Finally we introduce a factor 1/2, $M_i$ = $\frac{1}{2}\widehat{M}_i$, for model consistency with experimental data in the dilute limit (compare Appendix B):
\begin{eqnarray}
\frac{\partial x_j}{\partial t}&=&\frac{1}{2}\nabla \sum_{\substack{i=1 \\ i \neq j}}^n\bigg[x_ix_j\bigg(x_iM_j+x_jM_i+ \nonumber \\
&& \sum_{\substack{k=1\\k\neq j\\k\neq i}}^{n}x_k(M_i+M_j-M_k)\bigg)\nabla(\mu_j-\mu_i)\bigg]
\label{Factor 1/2}
\end{eqnarray}

\section*{Appendix B - Dilute solution limit of pair-wise diffusion model}
\renewcommand{\theequation}{B\arabic{equation}}
\setcounter{equation}{0}

For simplification the Gibbs energy is given by an ideal solution model:
\begin{align}
G^{\text{ideal}}&=\sum_{i=1}^{n}x_iG_i(T)-TS^{\text{ideal}}\\
&=\sum_{i=1}^{n}x_iG_i(T)+RT\sum_{i=1}^{n}x_i\ln{(x_i)}
\end{align}
The derivatives of the chemical potentials with respect to site fractions are
given as $\frac{\partial \mu_j}{\partial x_j}=\frac{RT}{x_j}$ and $\frac{\partial \mu_j}{\partial x_k}=-\frac{RT}{x_j}$.\\
Rewriting the diffusion equation depending on the concentration gradient:
\begin{eqnarray}
\frac{\partial x_j}{\partial t}&=&\frac{1}{2}\nabla\sum_{\substack{i=1 \\ i \neq j}}^n\bigg[\bigg(\sum_{\substack{k=1\\k\neq j\\k\neq i}}^{n}x_ix_jx_k(M_i+M_j-M_k)+ \nonumber \\
&& x_ix_j(x_iM_j+x_jM_i)\bigg)\sum_{l=1}^{n}\bigg(\frac{\partial \mu_j}{\partial x_l}-\frac{\partial \mu_i}{\partial x_l}\bigg)\nabla x_l\bigg]
\end{eqnarray}
Replacing the derivatives of the chemical potential:
\begin{eqnarray}
\frac{\partial x_j}{\partial t}&=&\frac{1}{2}\nabla\sum_{\substack{i=1 \\ i \neq j}}^n\bigg[\bigg(\sum_{\substack{k=1\\k\neq j\\k\neq i}}^{n}x_ix_jx_k(M_i+M_j-M_k)+ \nonumber \\
&& x_ix_j(x_iM_j+x_jM_i)\bigg) \times \nonumber \\
&& \bigg(\sum_{\substack{l=1\\l\neq i \\ l \neq j}}^{n}\bigg(-\frac{RT}{x_j}+\frac{RT}{x_i}\bigg)\nabla x_l
+ \nonumber \\
&&\bigg(-\frac{RT}{x_j}-\frac{RT}{x_i}\bigg)\nabla x_i+ \nonumber \\
&&\bigg(\frac{RT}{x_j}+\frac{RT}{x_i}\bigg)\nabla x_j\bigg)\bigg]
\end{eqnarray}
 Replace the gradient of $x_i$: $\nabla x_i = -\sum_{\substack{m=1\\m\neq i}}^{n}\nabla x_m$ and rewrite the equation:
\begin{eqnarray}
\frac{\partial x_j}{\partial t}&=&RT\nabla\sum_{\substack{i=1 \\ i \neq j}}\bigg[
\bigg(\sum_{\substack{k=1\\k\neq j}}^{n}x_jx_kM_j-\sum_{\substack{k=1\\k\neq j}}^{n}x_jx_kM_k+x_jM_i\bigg) 
\times \nonumber \\
&&\sum_{\substack{l=1\\l\neq i}}^{n}\nabla x_l
+\bigg(x_iM_j-x_jx_iM_j+x_jx_iM_i\bigg)\nabla x_j\bigg]
\end{eqnarray}
In the dilute limit: $x_j \rightarrow 0$:
\begin{align}
\frac{\partial x_j}{\partial t}&=RT\nabla\sum_{\substack{i=1 \\ i \neq j}}x_iM_j\nabla x_j
\end{align}
\begin{align}
\frac{\partial x_j}{\partial t}&=RT\nabla M_j\nabla x_j
\end{align}
Note that this convention is different to the DICTRA convention by a factor of 2 (compare Eq.~\ref{Factor 1/2}).

\section*{Appendix C - Data repositories}
The composition dependent atomic mobilities (except for Ni) in the MTIC-Lin data
repository are given as: (in m$^2$Js$^{-1}$mol$^{-1}$ at $T=1373$ K)
\begin{align*}
M_{\text{Co}} &= -7.90\times10^{-19} + 3.03\times10^{-19}\cdot x_{\text{Co}}\\
M_{\text{Cr}} &= -14.9\times10^{-19} + 7.15\times10^{-19}\cdot x_{\text{Co}}\\
M_{\text{Fe}} &= -15.8\times10^{-19} + 6.05\times10^{-19}\cdot x_{\text{Co}}\\
M_{\text{Mn}} &= -45.2\times10^{-19} + 18.5\times10^{-19}\cdot x_{\text{Co}}\\
M_{\text{Ni}} &= 7.3\times10^{-20}
\end{align*}
These equations can be rewritten in the Redlich-Kister expansion compatible with
the DICTRA-notation: (in Jmol$^{-1}$)\\
Mobility of Co:
\begin{align*}
\text{MQ(FCC,CO:VA,0)} &= Q_{\text{Co}}^{\text{Co}} = -430713\\
\text{MQ(FCC,CR:VA,0)} &= Q_{\text{Co}}^{\text{Cr}} = -368291\\
\text{MQ(FCC,FE:VA,0)} &= Q_{\text{Co}}^{\text{Fe}} = -368291\\
\text{MQ(FCC,MN:VA,0)} &= Q_{\text{Co}}^{\text{Mn}} = -368291\\
\text{MQ(FCC,NI:VA,0)} &= Q_{\text{Co}}^{\text{Ni}} = -368291
\end{align*}
Mobility of Cr:
\begin{align*}
\text{MQ(FCC,CO:VA,0)} &= Q_{\text{Cr}}^{\text{Co}} = -360576\\
\text{MQ(FCC,CR:VA,0)} &= Q_{\text{Cr}}^{\text{Cr}} = -400765\\
\text{MQ(FCC,FE:VA,0)} &= Q_{\text{Cr}}^{\text{Fe}} = -360576\\
\text{MQ(FCC,MN:VA,0)} &= Q_{\text{Cr}}^{\text{Mn}} = -360576\\
\text{MQ(FCC,NI:VA,0)} &= Q_{\text{Cr}}^{\text{Ni}} = -360576
\end{align*}
Mobility of Fe:
\begin{align*}
\text{MQ(FCC,CO:VA,0)} &= Q_{\text{Fe}}^{\text{Co}} = -360546\\
\text{MQ(FCC,CR:VA,0)} &= Q_{\text{Fe}}^{\text{Cr}} = -360546\\
\text{MQ(FCC,FE:VA,0)} &= Q_{\text{Fe}}^{\text{Fe}} = -423196\\
\text{MQ(FCC,MN:VA,0)} &= Q_{\text{Fe}}^{\text{Mn}} = -360546\\
\text{MQ(FCC,NI:VA,0)} &= Q_{\text{Fe}}^{\text{Ni}} = -360546
\end{align*}
Mobility of Mn:
\begin{align*}
\text{MQ(FCC,CO:VA,0)} &= Q_{\text{Mn}}^{\text{Co}} = -348709\\
\text{MQ(FCC,CR:VA,0)} &= Q_{\text{Mn}}^{\text{Cr}} = -348709\\
\text{MQ(FCC,FE:VA,0)} &= Q_{\text{Mn}}^{\text{Fe}} = -348709\\
\text{MQ(FCC,MN:VA,0)} &= Q_{\text{Mn}}^{\text{Mn}} = -402756\\
\text{MQ(FCC,NI:VA,0)} &= Q_{\text{Mn}}^{\text{Ni}} = -348709
\end{align*}
Mobility of Ni:
\begin{align*}
\text{MQ(FCC,CO:VA,0)} &= Q_{\text{Ni}}^{\text{Co}} = -387994\\
\text{MQ(FCC,CR:VA,0)} &= Q_{\text{Ni}}^{\text{Cr}} = -387994\\
\text{MQ(FCC,FE:VA,0)} &= Q_{\text{Ni}}^{\text{Fe}} = -387994\\
\text{MQ(FCC,MN:VA,0)} &= Q_{\text{Ni}}^{\text{Mn}} = -387994\\
\text{MQ(FCC,NI:VA,0)} &= Q_{\text{Ni}}^{\text{Ni}} = -387994
\end{align*}


\begin{thebibliography}{}

\bibitem{Murty2014} B.S. Murty, J.W. Yeh, S. Ranganathan, High Entropy
Alloys. Elsevier, London (2014).
\bibitem{Yeh2004} J.W. Yeh, S.K. Chen, S.J. Lin, J.Y. Gan, T.S. Chin, T.T. Shun,
C.H. Tsau, S.Y. Chang, Nanostructured high-entropy alloys with multiple
principal elements: novel alloy design concepts and outcomes, Adv. Eng. Mater
6 (2004) 299-303.
\bibitem{Zhang2014} F. Zhang, C. Zhang, S.K. Chen, J. Zhu, W.S. Cao, U.R.
Kattner, An understanding of high entropy alloys from phase diagram
calculations. Calphad 45 (2014) 1-10.
\bibitem{Ma2015} D. Ma, B. Grabowski, F. K\"ormann, J. Neugebauer, D. Raabe,
Ab initio thermodynamics of the CoCrFeMnNi high entropy alloy: importance of
entropy contributions beyond the configurational one, Acta Mater 100 (2015) 
90-97.
\bibitem{Schuh2015} B. Schuh, F. Mendez-Martin, B. V\"olker, E.P. George, H.
Clemens, R. Pippan, A. Hohenwarter, Mechanical properties,
microstructure and thermal stability of a nanocrystalline CoCrFeMnNi high-entropy alloy after
severe plastic deformation, Acta Mater 96 (2015) 258-268.
\bibitem{Otto2016} F. Otto, A. Dlouhý, K.G. Pradeep, M. Kuběnová, D.
Raabe, G.
Eggeler, E.P. George, Decomposition of the single-phase high-entropy
alloy CrMnFeCoNi after prolonged anneals at intermediate temperatures, Acta
Mater 112, (2016) 40-52.
\bibitem{Guo2016} N.N. Guo, L. Wang, L.S. Luo, X.Z. Li, R.R. Chen, Y.Q. Su, J.J.
Guo, H.Z. Fu, Hot deformation characteristics and dynamic
recrystallization of the MoNbHfZrTi refractory high-entropy alloy, Mater. Sci.
Eng. A 651 (2016) 698-707.
\bibitem{Chen2016} H. Chen, A. Kauffmann, B. Gorr, D. Schliephake, C. 
Seem\"uller, J.N. Wagner, H.-J. Christ, M. Heilmaier, Microstructure and
mechanical properties at elevated temperatures of a new Al-containing refractory
high-entropy alloy Nb-Mo-Cr-Ti-Al, J. Alloys Compd. 661 (2016) 206-215.
\bibitem{Lee2016} D.H. Lee, M.Y. Seok, Y. Zhai, I.C. Choi, J. He, Z. Lu, J.Y.
Suh, U. Ramamurty, M. Kawasaki, T.G. Langdon, J.I. Jang, Spherical
nanoindentation creep behavior of nanocrystalline and coarse-grained CoCrFeMnNi
high-entropy alloys, Acta Mater 109 (2016) 314-322.
\bibitem{Zhang2016} L. Zhang, P. Yu, H. Cheng, H. Zhang, H. Diao, Y. Shi, B.
Chen, P. Chen, R. Feng, J. Bai, Q. Jing, M. Ma, P.K. Liaw, G. Li, R. Liu, 
Nanoindentation creep behavior of an Al0.3CoCrFeNi high-entropy alloy,
Metall. Mater. Trans. A (2016) 1-5.
\bibitem{Ma2016} Y. Ma, Y.H. Feng, T.T. Debela, G.J. Peng, T.H. Zhang, 
Nanoindentation study on the creep characteristics of high-entropy alloy
films: fcc versus bcc structures, Int. J. Refract. Met. H. 54 (2016) 395-400.
\bibitem{Cao2016} T. Cao, J. Shang, J. Zhao, C. Cheng, R. Wang, H. Wang,
The influence of Al elements on the structure and the creep behavior of AlxCoCrFeNi
high entropy alloys, Mater. Lett. 164 (2016) 344-347.
\bibitem{Kai2016} W. Kai, C.C. Li, F.P. Cheng, K.P. Chu, R.T. Huang, L.W. Tsay,
J.J. Kai, The oxidation behavior of an equimolar FeCoNiCrMn high-entropy
alloy at 950 $^{\circ}$C in various oxygen-containing atmospheres, Corros. Sci.
108 (2016) 209-214.
\bibitem{Laplanche2016} G. Laplanche, U.F. Volkert, G. Eggeler, E.P. George,
 Oxidation behavior of the CrMnFeCoNi high-entropy alloy, Oxid. Met.
85 (2016) 629-645.
\bibitem{Holcomb2015} G.R. Holcomb, J. Tylczak, C. Carney, Oxidation of
CoCrFeMnNi high entropy alloys, JOM 67 (2015) 2326-2339.
\bibitem{Shaginyan2016} R.A. Shaginyan, N.A. Krapivka, S.A. Firstov, N.I.
Danilenko, I.V. Serdyuk, Superhard vacuum coatings based on high-entropy
alloys, Powder Metall. Met. C+ 54 (2016) 725-730.
\bibitem{Pickering2016} E.J. Pickering, N.G. Jones, High-entropy alloys:
a critical assessment of their founding principles and future prospects, Int.
Mater. Rev. (2016) 1-20.
\bibitem{cors}D.B. Miracle: High-entropy alloys, A current evaluation of 
founding ideas and core effects and exploring "nonlinear alloys", JOM
69 (2017) 2130-2136.
\bibitem{Praveen2016} S. Praveen, J. Basu, S, Kashyap, R.S. Kottada,
 Exceptional resistance to grain growth in nanocrystalline CoCrFeNi high entropy
alloy at high homologous temperatures, J. Alloys Compd. 662 (2016) 361-367.
\bibitem{Tsai2013} K.Y. Tsai, M.H. Tsai, J.W. Yeh, Sluggish diffusion in
Co-Cr-Fe-Mn-Ni high-entropy alloys, Acta Mater 61 (2013) 4887-4897.
\bibitem{Kulkarni2015} K. Kulkarni, G.P.S. Chauhan, Investigations of
quaternary interdiffusion in a constituent system of high entropy alloys, AIP
Adv. 5 (2015) 097162.
\bibitem{Dabrowa2016} J. Dabrowa, W. Kucza, G. Cieslak, T. Kulik, M.
Danielewski, J.W. Yeh, Interdiffusion in the FCC-structured
Al-Co-Cr-Fe-Ni high entropy alloys: experimental studies and numerical
simulations, J. Alloys Compd. 674 (2016) 455-462.
\bibitem{Vaidya2016}M. Vaidya, S. Trubel, B.S. Murty, G. Wilde, 
S.V. Divinski, Ni tracer diffusion in CoCrFeNi and CoCrFeMnNi high
entropy alloys, JALCOM 688 (2016) 994-1001.
\bibitem{Vaidya2017}M. Vaidya, K.G. Pradeep, B.S. Murty, G. Wilde, S.V. 
Divinski, Radioactive isotopes reveal a non sluggish kinetics of grain
boundary diffusion in high entropy alloys, Scientific Reports 7 (2017) 12273.
\bibitem{Vaidya2018}M. Vaidya, K.G. Pradeep, B.S. Murty, G. Wilde, S.V. 
Divinski, Bulk tracer diffusion in CoCrFeNi and CoCrFeMnNi high entropy
alloys, Acta Mater 146 (2018) 211-224.
\bibitem{Gaertner2018}D. Gaertner, J. Kottke, Y. Chumlyakov, G. Wilde, S.V.
Divinski, Tracer diffusion in single crystalline CoCrFeNi and
CoCrFeMnNi high entropy alloys, JMR (2018) \emph{in press}. 
\bibitem{Boettinger2016} W. J. Boettinger, J.E. Guyer, C.E. Campbell, G.B.
McFadden, Computation of the Kirkendall velocity and displacement fields in a
one-dimensional binary diffusion couple with a moving interface, Proc. R. Soc. A 463 (2007) 3347-3373.
\bibitem{Chen2018} Q. Chen,  A. Engstr\"om, J. {\AA}gren, On Negative Diagonal Elements in the Diffusion Coefficient Matrix of Multicomponent Systems, J. Phase Equilib. Diffus. 39 (2018) 592-596.
\bibitem{Paul2013} A. Paul, A pseudobinary approach to study
interdiffusion and the Kirkendall effect in multicomponent systems, Philos.
Mag. 93 (2013) 2297-2315.
\bibitem{Murch2017}T.R. Paul, I.V. Belova, G.E. Murch, Analysis of
diffusion in high entropy alloys, Mater. Chem. Phys. 210 (2017) 301-308.
\bibitem{Review}S.V. Divinski, A. Pokoev, N. Eesakkiraja, A. Paul, A mystery of
'sluggish diffusion' in high-entropy alloys: the truth or a myth?, Diffusion Foundations (2018) accepted.
\bibitem{Vaidya2018S}M. Vaidya, M. Muralikrishna, S.V. Divinski, B.S. Murty,
Non-sluggish interdiffusion kinetics in CoCrFeNi and CoCrFeMnNi high entropy
alloys at elevated temperature, Scripta Mater 157 (2018) 81-85.
\bibitem{Agren1982} J. {\AA}gren, Diffusion in phases with several
components and sublattices, J. Phys. Chem. Solids 43 (1982) 5.
\bibitem{Andersson1992}  J.-O. Andersson, J. {\AA}gren, Models for
numerical treatment of multicomponent diffusion in simple phases,
J. Appl. Phys. 72 (1992) 4.
\bibitem{Borgenstam2000}A. Borgenstam, A. Engstr\"om, L. H\"oglund, J.
{\AA}gren, DICTRA, a tool for simulation of diffusional transformations in
alloys, J. Phase Equilibr. 21 (2009) 269.
\bibitem{Zhang2015} L. Zhang, M. Stratmann, Y. Du, B. Sundmann, I. Steinbach, Incorporating the CALPHAD sublattice approach of ordering into the phase-field model with finite interface dissipation, Acta Mater 88 (2015) 156-159.
\bibitem{Lukas2010} H. L. Lukas, S.G. Fries, B. Sundman, 
Computational Thermodynamics. The Calphad Method,
Cambridge University Press, Cambridge (2010).
\bibitem{Campbell2001} C.E. Campbell, W.J. Boettinger, U.R. Kattner, Development
of a diffusion mobility database for Ni-base superalloys, Acta Mat. 50 (2002).
\bibitem{TCHEA3} Thermo-Calc Software TCS High Entropy Alloys Database
Version 3.
\bibitem{Haase2017} C. Haase, F. Tang, M. B. Wilms, A. Weisheit, B.Hallstedt,
 Combining thermodynamic modeling and 3D printing of elemental powder
blends for high-throughput investigation of high-entropy alloys - Towards rapid
alloy screening and design, Mater. Sci. Eng., A 688 (2017) 180-189.
\bibitem{TCHEA_dat}  Thermo-Calc Software TCS High Entropy Alloy Database and
Thermo-Calc Software TCS High Entropy Alloy Mobility Database.
\bibitem{TCNI_dat} Thermo-Calc Software, TCNI Ni-based Superalloys Database
Version 8 (accessed 4 June 2018).
\bibitem{Lemmer1955}H.R. Lemmer, O.J.A. Segaert, M.A. Grace,
The decay of Cobalt 57, Proc. Phys. Soc. A 68 (1955) 701-708.
\bibitem{Ofer1957} S. Ofer, R. Wiener, Decay of Cr 51, Phys. Rev.
107 (1957) 1639-1641.
\bibitem{Heath1960} R.L. Heath, C.W. Reich, D.G. Proctor, Decay of
45-Day Fe 59, Phys. Rev. 118 (1960) 1082.
\bibitem{Lederer1978} C.M. Lederer, V.S. Shirley, Table of Isotopes, 7th                   
ed, Wiley, New York (1978).
\bibitem{Mehrer2007} H. Mehrer, Diffusion in Solids: Fundamentals,
Methods, Materials, Diffusion-controlled Processes, Springer, Berlin (2007).
\bibitem{Belova2015} I. Belova, Y. Sohn, G. Murch, Measurement of
tracer diffusion coefficients in an interdiffusion context for multicomponent
alloys, Phil. Mag. Let. 95 (2015) 416-424.
\bibitem{Sauer1962} F. Sauer, V. Freise, Diffusion in binary alloys
with volume change, Z. Elektrochem. 66 (1962) 353.
\bibitem{Redlich1948} O. Redlich, A.T. Kister, Algebraic representation
of thermodynamic properties and the classification of solutions, Ind. and Eng.
Chem. 40 (1948) 345-348.
\bibitem{Divinski2001} S.V. Divinski, M. Lohmann, Ch. Herzig, Ag grain bondary
diffusion and segregation in Cu:Measurements in the types B and C diffusion
regimes, Acta Mater 49 (2001) 249-261.
\bibitem{Kubo1966} R. Kubo, The fluctuation-dissipation theorem, Rep. Prog. Phys. 29 (1966) 255-284.
\bibitem{TCFE_dat} Thermo-Calc Software, TCFE Steels/Fe-alloys Database Version
9 (accessed 8 December 2017). 


\end{thebibliography}
\end{document}